\documentclass[a4paper,fleqn]{cas-sc}

\usepackage{lipsum}
\usepackage{graphicx}
\usepackage{nth}
\usepackage{multirow}
\usepackage{amsmath}
\usepackage{color}
\usepackage{makecell}

\definecolor{cardinal}{cmyk}{0,1,0.63,0.29}
\definecolor{orange}{cmyk}{0, 0.5333, 0.8, 0}

\newcommand{\eg}{\textit{e.g.,\ }}
\newcommand{\etc}{\textit{etc.\@}}

\newcommand{\ie}{\textit{i.e.,\ }}

\newcommand{\cmmnt}[1]{}
\newcommand{\topic}[1]{{\color{blue}{\emph{#1}}}}
\renewcommand{\topic}[1]{}

\usepackage[authoryear,longnamesfirst]{natbib}
\let\printorcid\relax

\begin{document}

\let\WriteBookmarks\relax
\def\floatpagepagefraction{1}
\def\textpagefraction{.001}
\shorttitle{Unsupervised Speech Representation Learning for Behavior Modeling using Triplet Enhanced Contextualized Networks}
\shortauthors{Haoqi Li et~al.}

\title [mode = title]{Unsupervised Speech Representation Learning for Behavior Modeling using Triplet Enhanced Contextualized Networks} 
\author[1]{Haoqi Li}
\ead{haoqili@usc.edu}

\author[2]{Brian Baucom}
\ead{brian.baucom@psych.utah.edu}

\author[1]{Shrikanth Narayanan}
\ead{shri@ee.usc.edu}

\author[1]{Panayiotis Georgiou}
\ead{georgiou@ee.usc.edu}

\address[1]{Signal Analysis and Interpretation Laboratory (SAIL), University of Southern California, Los Angeles, USA}
\address[2]{Department of Psychology, University of Utah, Salt Lake City, USA}

\begin{abstract}
Speech encodes a wealth of information related to human behavior and has been used in a variety of automated behavior recognition tasks.
However, extracting behavioral information from speech remains challenging including due to inadequate training data resources stemming from the often low occurrence frequencies of specific behavioral patterns. 
Moreover, supervised behavioral modeling typically relies on domain-specific construct definitions and corresponding manually-annotated data, rendering generalizing across domains challenging. 
In this paper, we exploit the stationary properties of human behavior within an interaction and present a representation learning method to capture behavioral information from speech in an unsupervised way.
We hypothesize that nearby segments of speech share the same behavioral context and hence map onto similar underlying behavioral representations.
We present an encoder-decoder based Deep Contextualized Network (DCN) as well as a Triplet-Enhanced DCN (TE-DCN) framework to capture the behavioral context and derive a  manifold representation, where speech frames with similar behaviors are closer while frames of different behaviors maintain larger distances.
The models are trained on movie audio data and validated on diverse domains including on a couples therapy corpus and other publicly collected data (e.g., stand-up comedy).
With encouraging results, our proposed framework shows the feasibility of unsupervised learning within cross-domain behavioral modeling.
\end{abstract}


\maketitle

\section{Introduction}
\topic{what behavior analysis is.}
Human behavior refers to the way humans act and interact in response to a stimulus, internal or external. 
Understanding human behavior through observational study is one of the core methodologies in fields such as psychology and sociology \citep{margolin1998nuts}. 
Human behaviors encompass rich information: from emotional expression, processing, and regulation to the intricate dynamics of interactions, including the context and knowledge of interlocutors and their thinking and problem-solving intent \citep{li2019linking}. Furthermore, the behavioral constructs of interest are often dependent on the domain of interaction \citep{narayanan2013behavioral}.   
Hence characterization of human behavior usually requires domain-specific knowledge and adequate windows of observation. 
Notably, across psychological health science and practice \citep{Bone2017SignalProcessingandMachine} such as couple therapy \citep{christensen2004traditional},  suicide cognition evaluation \citep{bryan2014improving} and addiction counseling \citep{Xiao2015RatemytherapistAutomated}, this is exemplified in the definition and derivation of a variety of domain-specific behavior constructs (e.g., blame and affect patterns exhibited by partners, suicidal ideation of an individual at risk, and empathy expressed by a therapist in the respective aforementioned domains) to support specific subsequent plan of action. 

\topic{BSP via speech cues}
Human speech offers rich information about the mental state and traits of the talkers. 
Vocal cues, including speech and spoken language as well as nonverbal vocalizations and disfluency patterns, have been shown to be informationally relevant in the context of human behavior (\eg in marital interaction \citep{baucom2009prediction}, in motivational interviewing \citep{amrhein2003client, imel2014association, miller1993enhancing}). 
Many automatic computational approaches that support measurement, analysis, and modeling of human behaviors from speech have been investigated in affective computing \citep{Lee2005Towarddetectingemotionsin}, social signal processing \citep{vinciarelli2009social} and behavioral signal processing (BSP) \citep{narayanan2013behavioral}.

\topic{Challenges}
Automated behavior modeling from speech however remains a challenging domain. 
Behavior annotations used for (supervised) modeling are usually obtained from well-trained human annotators, in a process that is both complex and expensive. 
Moreover, the prevalence of many specific behaviors of interest in a given interaction inherently tend to be low.
As a result, the amount of annotated training data available for supervised behavior modeling are relatively small compared to other speech related training tasks.

In addition, behavior analyses tend to be guided by target domain needs. 
For example, in looking for markers of behavior change in addiction, therapists look for language which reflects changes of addictive habits \citep{baer2009agency}. 
In suicide prevention \citep{cummins2015review}, behavioral patterns related to reasons for living and emotional bonds are deemed relevant. 
Thus,  behavior models built with domain-specific constructs and data may not be directly and easily adaptable  across domains. 

\topic{motivation}
Recently, unsupervised and self-supervised learning \citep{latif2020deep, chen2020simple} have shown the benefits of using large amounts of unlabelled data to extract informative representations.
Given the low availability of annotated behavioral data sets, representation learning through unsupervised ways can provide a promising avenue for behavioral modeling. 
This becomes especially relevant where unlabelled or weakly-labelled speech is often the only available resource. 

In unsupervised representation learning, context information has been used for a range of applications \citep{goldberg2014word2vec, devlin2018bert}. For example, in Natural Language Processing (NLP), word and sentence embedding methods attempt to compress the shared structural information between neighboring words, phrases or sentences. Such compressed structural information, referred as the \emph{context}, resides at a longer scale than either of just two neighboring isolated words, phrases or sentences. 
In behavior analysis, context information is important. When a human attempts to evaluate behaviors, a large observation window is often employed to observe the context. We can assume that behavior remains relatively constant within a sufficiently long observation window. This matches annotation guidelines in the field of psychology where the minimum observation windows are usually set at around 30 seconds. This assumption also matches our empirical understanding of human behaviors. For example, a person (often the case in couples therapy interactions as well as daily life) can be sad during a conversation for a sufficiently long time despite different speech patterns or intonations throughout that temporal window.



In this paper, we describe unsupervised methods to extract behavior related representations from speech under the behavioral stationarity assumption.
In addition, we also employ metric learning techniques to improve representation learning directly in the behavior related manifold space. 
We investigate whether out-of-domain data corpora can be employed for behavior representation learning and, for quantification and analysis of target behavioral constructs.
Moreover, we show the proposed unsupervised model can provide domain experts with dynamic behavior change trajectories, which can be helpful in facilitating the annotation process and highlight salient behavior regions.
To evaluate our proposed methods, we use a couple therapy dataset comprising audio recordings of problem solving interactions as well as speech files from a variety of application domains such as talk shows to show the similarity in the learned behavior manifolds. 


\section{Related work and motivation}
\label{sec:relatedwork}
The human speech audio includes information about the state and trait of the talkers ranging at varying levels of linguistic scales, \eg  phonemic, prosodic, and discourse, to the level of the larger socio-emotional communication context.

Traditional supervised behavior recognition systems mainly depend on two aspects: one is the representative feature of the target behavior and, the other is the choice of the classification model. 
To capture the vocal cues for behavior recognition, traditional computational approaches \citep{schuller2009interspeech, black2013toward, xia2015dynamic,li2016sparsely,nasir2017predicting} use a range of hand-crafted low-level descriptors (LLDs) (\eg f0, intensity, MFCCs (Mel-Frequency Cepstral Coefficients) \etc) with statistical functionals (\eg mean, median, standard deviation, \etc) to represent segment- or utterance-level features.
Based on these raw acoustic LLDs and their functionals, classifiers such as Support Vector Machines (SVM), k-Nearest Neighbors (kNN) and Hidden Markov Models (HMM) \textit{etc.} have been employed \citep{zeng2009survey, hu2007gmm, schuller2004speech, el2011survey, xia2015dynamic}. 

Over the last few years, many affect and behavior recognition systems have employed Deep Neural Network (DNN) models to extract intermediate representations \citep{han2014speech, li2013hybrid,li2016sparsely}. Further, sequential models \citep{lee2015high, li2019linking} have been used to account for the context effect. 
However, the success of DNN models heavily relies on the availability of large-scale datasets. 
A large amount of training data with annotated labels are usually unavailable in the human behavioral modeling related domains, which largely inhibits the use of DNN based supervised frameworks in behavioral modeling tasks \citep{li2016sparsely}.   


Different from supervised approaches, in this paper, we focus on context-rich techniques for extracting behavior representations in an unsupervised manner. 
Contextual information has played a significant role in unsupervised representation learning for a range of applications. 
For example, in NLP, contextual information is employed to generate general word or sentence embeddings (\eg Word2Vec \citep{mikolov2013distributed, mikolov2013efficient, goldberg2014word2vec}, BERT \citep{devlin2018bert} \etc) for downstream tasks.
In speech representation learning \citep{latif2020deep}, unsupervised techniques such as autoregressive modeling \citep{chung2019unsupervised, chung2020generative, chung2020improved} and self-supervised modeling \citep{milde2018unspeech, tagliasacchi2019self, pascual2019learning} employ temporal context information for extracting speech representation. 
In our prior behavior modeling work, an unsupervised representative learning framework was proposed \citep{haoqibeh2vec},  which showed the promise of learning behavior representations based on the behavior stationarity hypothesis that nearby segments of speech share the same behavioral context. A similar framing was used by \citet{nasir2018towards} to evaluate interpersonal entrainment through an unsupervised turn-level distance measure . 

In addition, metric learning is often employed to directly learn representations with an appropriate distance metric. 
For instance, siamese networks \citep{bromley1994signature} and triplet networks \citep{hoffer2015deep} are neural networks suitable for direct representation learning by minimizing the contrastive loss or triplet loss calculated in the latent embedding space. These techniques have shown promising results in face verification and identification \citep{schroff2015facenet} as well as in speech tasks such as speaker diarization and verification \citep{jati2019neural, song2018triplet}. 

The goal of this work is to identify, in an unsupervised manner, a latent manifold in which behavior characteristics are retained while other unrelated information are minimized. 
We believe the unsupervised representation learning under the behavioral stationarity assumption can take advantage of diverse out-of-domain datasets for improving behavioral modeling. 

\section{Unsupervised speech representation learning for human behavior modeling}
\label{sec:method}

We present two frameworks for unsupervised behavior modeling. The first one is the \textit{Deep Contextualized Network} (DCN) initially introduced by \citet{haoqibeh2vec}, and the second is a new hybrid approach enhanced by a triplet loss, referred to as \textit{Triplet Enhanced Deep Contextualized Network} (TE-DCN).
The overarching goal is to build a function that can map behavior related information from raw acoustic features into the behavioral manifold, where similar behaviors can be clustered closer than they are in the original acoustic feature space, while distinct behavior types can maintain larger distances between one another.

\subsection{Behavioral stationarity assumption}
Toward designing the unsupervised modeling, we wish to invoke some domain knowledge about human behaviors. 
An important observation is that complex human behaviors often manifest over longer time scales, and remain relatively constant within a sufficiently long temporal window;   in fact, one needs a sufficiently long observation time for facilitating human annotation of target behavioral constructs (\eg ranging from 30 seconds to 10 minutes \citep{heyman2004rapid,heavey2002couples}). For example, in couple therapy, interaction behaviors associated with constructs such as sadness and blame can last over several conversational exchanges.

Based on these observations, we make the \textit{behavior stationarity assumption}: Human behaviors are deemed to remain constant within a sufficiently long window (\ie behavior  stationary region).
This means that by observing target behaviors within a desired long observation window (\eg 30 seconds), it is likely that the same or similar behavioral states are observed.

\subsection{Deep Contextualized Network}
\begin{figure}[ht]
\centering
\includegraphics[width=0.75\textwidth]{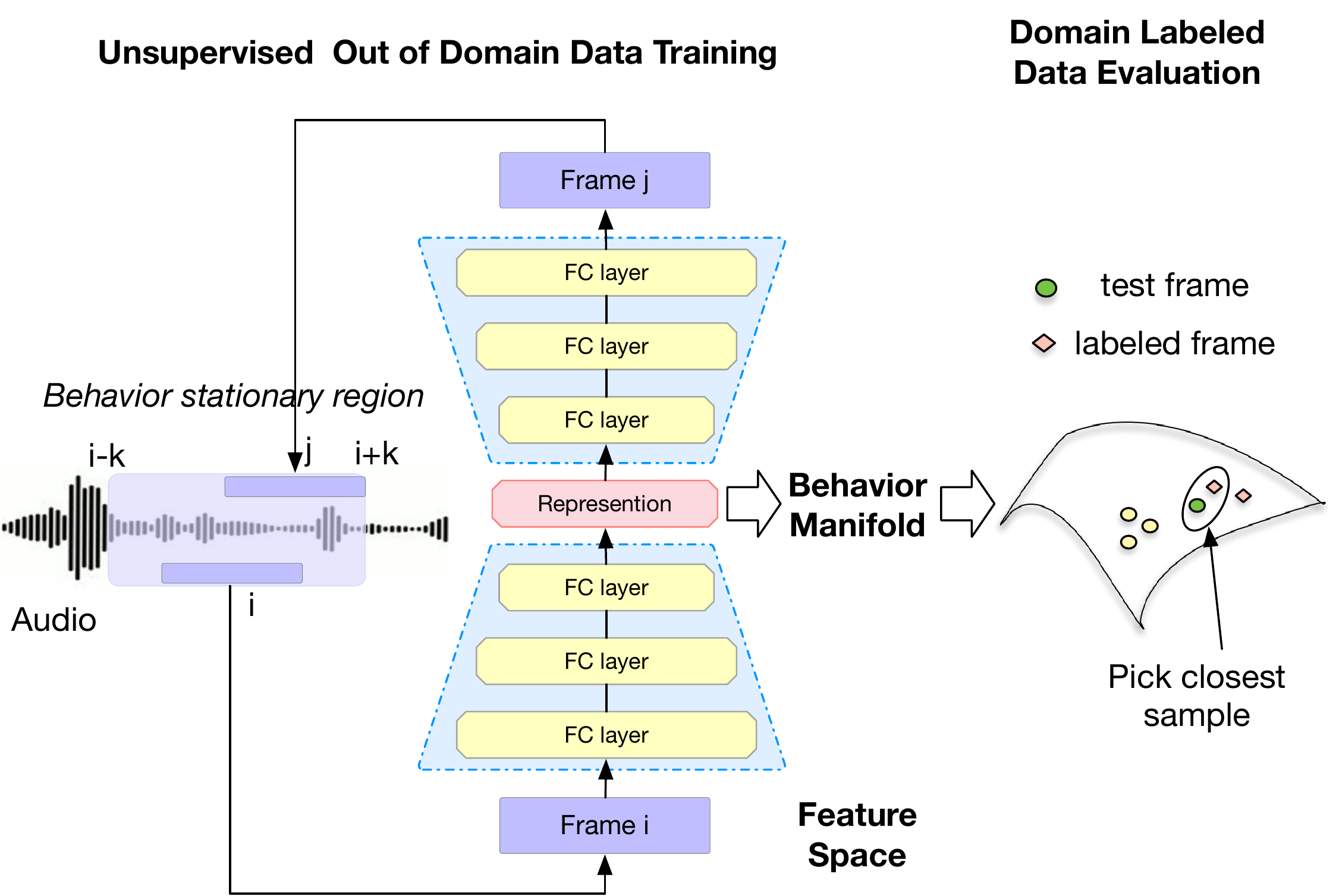}
\caption{Behavior representation learning framework via the DCN model. 
During training, the model encodes neighboring frames with a DCN. The choice of features promotes the behavior information as the extracted common information in the behavior manifold. 
During evaluation, similarity comparison is made by calculating distance within the behavior manifold.}
\label{fig:fig_DCN}
\end{figure}

The Deep Contextualized Network (DCN) has an encoder-decoder structure,  
similar to an autoencoder. But in contrast, rather than just training to reconstruct the input itself, the proposed DCN model is trained to reconstruct neighboring frames sharing the same behavioral context. The overall framework is shown in Figure \ref{fig:fig_DCN}.

As shown in Figure 1, the input can be one frame of the acoustic features within a training audio session. We name it $x_i$, where $i$ refers to the \textit{i\textsuperscript{th}} frame within one feature frame sequence. The reconstruction frame $x_j$ is selected from \textit{i-k} to \textit{i+k} excluding the \textit{i\textsuperscript{th}} frame, where \textit{k} is the maximum sampling shift size within the behavior stationary region, in which we assume the behavioral context to remain constant. 
During the training, we optimize the network to minimize the reconstruction loss:
\begin{equation}\label{eq:reconstruct_loss}
\begin{aligned}
\mathcal{L}_{DCN}(x_i, x_j) &= \sum_{(x_i, x_j)\in\mathcal{D}} \left \| f_{DCN}(x_i) - x_j \right \|_2^2
&=\sum_{(x_i, x_j)\in\mathcal{D}} \left \| \widehat{x_i} - x_j \right \|_2^2
\end{aligned}
\end{equation}
where the training dataset $\mathcal{D}$ consists of input tuples ($x_i, x_j$), and $\widehat{x_i}$ is the output of DCN. 

The representation from the hidden layers of DCN compresses the shared information between input and output. 
Once we input behavior-relevant acoustic features into the DCN, the trained encoder can be regarded as a feature extractor for obtaining the shared information between input and output. The choice of features and the model's structure can promote behavior as common information.
Assuming the adequacy of the behavioral stationarity assumption and the behavioral information contained in the input features, the model will ensure bottleneck embedding features that are relevant to the relatively constant factors, \ie the behavior related features. 

After training, the hidden bottleneck layer's output is used as the behavior representation for evaluation. 
Following which, similarity comparison can be made using choice distance metrics. For example, after learning the manifold on unsupervised data, a test sample can be compared with all known samples in the manifold space, and the closet match can be selected. More details of the evaluation can be found in the section \ref{subsec:eval_method}.

\subsection{Triplet Enhanced Deep Contextualized Network}
In this section, we introduce the Triplet Enhanced Deep Contextualized Network (TE-DCN), in which we use metric learning techniques to improve the performance of DCN. 

Metric learning aims to find an input-output mapping function over a vector space and is explicitly trained to build distance metrics among vectors.
Triplet loss enables neural networks to keep the embeddings belonging to the same class close to each other, while moving embeddings with different classes far apart. It is used for representation learning by direct optimization in the latent embedding space.
Suppose the training dataset $\mathcal{D}$ consists of input tuples ($x_a, x_p, x_n$): an anchor $x_a$, a positive sample $x_p$ which belongs to the same class as the anchor and a negative sample $x_n$ from a different class. 
The corresponding embedding ($e_a, e_p, e_n$) is generated by neural networks, and the model is trained to minimize the following loss function:

\begin{equation}\label{eq:triplet_loss}
\mathcal{L}_{triplet}(x_a, x_p, x_n)
= \sum_{(x_a, x_p, x_n)\in\mathcal{D}} \max[0,m+D(e_a, e_p)-D(e_a, e_n)]
\end{equation}
where $D(\cdot,\cdot)$ denotes the distance metric and $m$ is the parameter of margin value.
This objective function aims to ensure that, in the embedding space, the anchor sample is closer to the positive sample than it is to the negative sample by at least a margin $m$.

\begin{figure}[ht]
\centering
\includegraphics[width=0.58\textwidth]{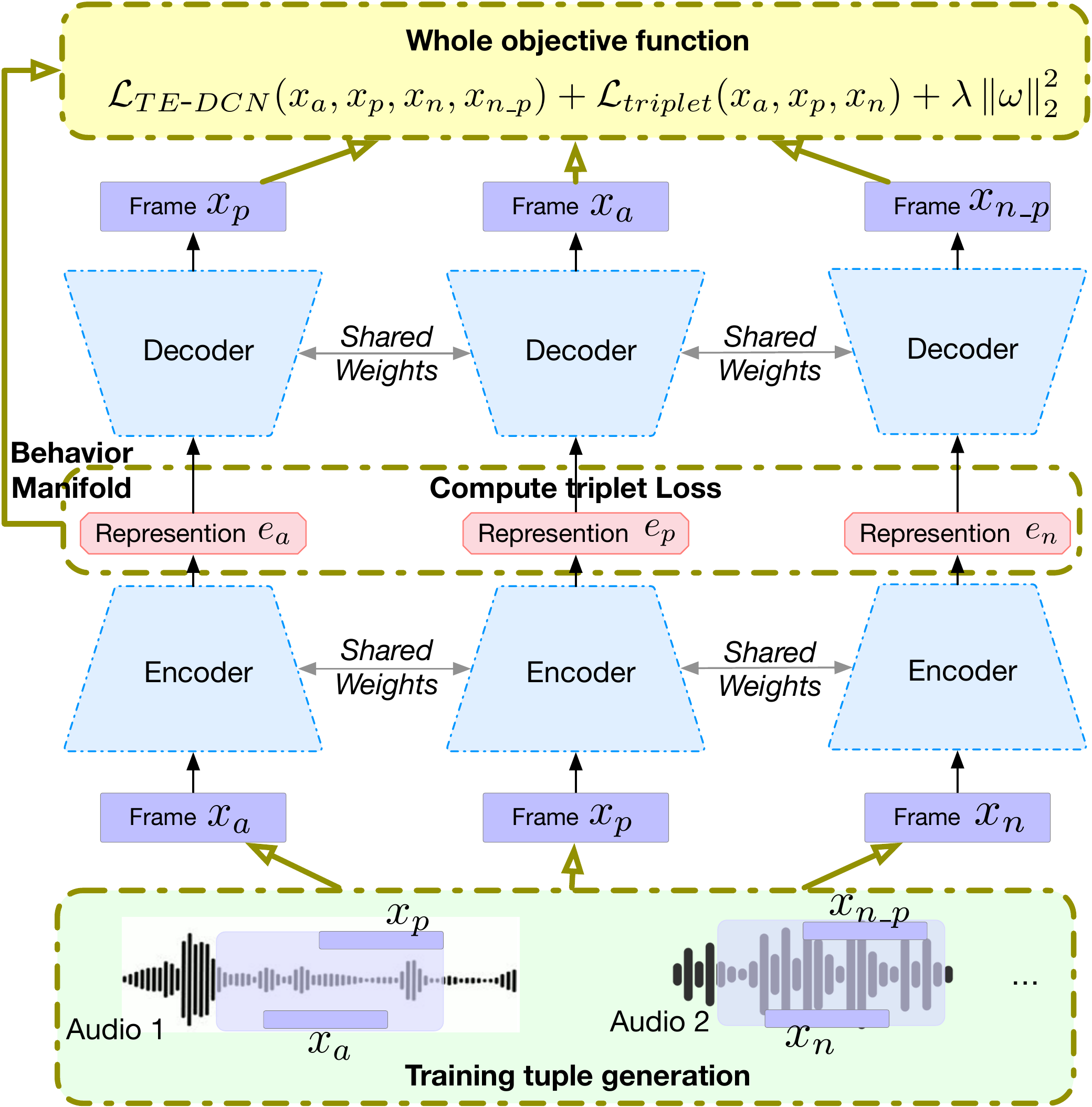}
\caption{Behavior representation learning framework via the TE-DCN model. The model has a triplet structure with shared weights. Audio 1 and 2 can be two temporally-distant regions or two different files. 
During training, the model is optimized to minimize both reconstruction loss and triplet loss. 
During evaluation, representation similarity comparison is performed within the behavior manifold. }
\label{fig:fig_TEDCN}
\end{figure}

The architecture of the proposed TE-DCN is shown in Figure \ref{fig:fig_TEDCN}.
The added triplet loss is motivated by a similar idea, in which we want to keep shared behavioral information between neighboring frames while disregarding other nuisance factors (with respect to the target behavioral construct), such as speaker and channel information. 
The DCN is only trained by frames with behavioral similarity in the stationary region, while the triplet loss requires frames from difference regions, which likely have different behaviors as well as distinct acoustic and speaker information. 

The model takes tuples of four elements ($x_a, x_p, x_n, x_{n\_p}$) for training, where ($x_a, x_p$) are two frames within a behavioral observation window (assuming stationarity within it). 
Let ($x_n, x_{n\_p}$) be another frame pair selected from a temporally-distant region, and is potentially more likely to contain a different target behavior.
Given this tuple, we have the input-output pair for our TE-DCN model: ($x_a, x_p, x_n$) and ($x_p, x_a, x_{n\_p}$). 
Each branch of the model can be regarded as a DCN framework with shared parameters to reconstruct context frames. 
In addition, the model also employs the triplet loss to directly optimize the intermediate embedding (i.e., the behavior manifold space).

In the TE-DCN model, we simultaneously optimize two objective functions (\ref{eq:reconstruct_loss}) and (\ref{eq:triplet_loss}) jointly.
Thus, the overall loss function is:

\begin{equation}\label{eq:total_loss}
\begin{aligned}
\mathcal{L}_{total}(x_a, x_p, x_n, x_{n\_p})
 =  \mathcal{L}_{TE\mbox{-}DCN}(x_a, x_p, x_n, x_{n\_p}) 
 +  \mathcal{L}_{triplet}(x_a, x_p, x_n) + \lambda  \left \| \omega \right \|_2^2
\end{aligned}
\end{equation}

where the reconstruction loss $\mathcal{L}_{TE\mbox{-}DCN}$ in Equation (\ref{eq:total_loss}) is defined as:

\begin{equation}\label{eq:3_reconstruction_loss}
\begin{aligned}
\mathcal{L}_{TE\mbox{-}DCN}(x_a, x_p, x_n, x_{n\_p}) =  \mathcal{L}_{DCN}(x_a, x_p) +  \mathcal{L}_{DCN}(x_p, x_a)  +\mathcal{L}_{DCN}(x_n, x_{n\_p})
\end{aligned}
\end{equation}

Since the model can take advantage of practically available (potentially) unlimited amount of unlabelled corpora, to prevent ovefitting in the training domain, as shown in Equation (\ref{eq:total_loss}), we amend the objective function with an $L_2$ regularization term. 

The encoder of the model tries to map the frame's acoustic features to a ``behavioral manifold''. On the one hand, the neighboring frames are trained to cluster together while on the other, frames from different regions are trained to be farther away in the representation space.

We propose the TE-DCN model to provide improvement to the DCN model in the following aspects:
\begin{enumerate}
\item \textit{Introduce discriminative information.} 
  
The metric learning is employed in the TE-DCN, which enables the model not only capture the behavioral contextual information within neighboring frames but also preserve the discriminative information by imposing triplet constraints.
By adding the triplet loss, the model not only reconstructs the frames pairs within behavioral stationary regions, but also uses ``negative" frames to reduce nuisance factor effects, such as speaker characteristics and channel information, in the behavior manifold construction.

\item \textit{Select the behavioral embedding layer explicitly.}

In DCN, the smallest bottleneck layer's output embedding is used as the behavior representation for evaluation. 
However, it is not guaranteed that this is the optimal choice among all the hidden layers.
In TE-DCN, though we choose the same smallest bottleneck embedding, we further use a triplet loss to directly add specific constraints and optimizations on that selected embedding.  Compared with other hidden embeddings, the selected embedding is directly optimized using both metric learning techniques and contextual information,  which promotes the extraction of targeted behavioral information on the explicit layer under the stationary assumption.  

\item  \textit{Uniformize the distance metric for training and evaluation.}

After learning from unlabelled data, during testing, we use distance metrics to evaluate the similarity within the behavioral manifold. However, the choice of distance measures is not specified for the DCN model. While in TE-DCN, the distance metric is unified in both training and testing stages, which can potentially reduce the uncertainty introduced by the selection of distance metric during evaluation.

\end{enumerate}

\section{Datasets}
\label{sec:dataset}
For the unsupervised training process, the training data should be easily acquired, and should include rich behavioral content and diverse set of conversations as much as possible.
In this work, we collected around 400 hours of audio from 225 movies\footnote{The list of collected movies can be found at github.com/haoqi/beh2vec}. 
Many of the selected movies include rich and diverse affective content reflecting a range of behaviors.
This training corpus is treated at the generic data set outside the target behavioral modeling domains. 
In previous behavior modeling corpora \citep{chakravarthula2019predicting,chakravarthula2019automatic, li2019linking}, there are at most 90 hours of original recording speech. Thus, compared to in-domain supervised behavior modeling tasks, we have a significantly larger amount of training data. 
We will use it to show the feasibility of proposed model within cross-domain behavioral modeling.

\subsection{Evaluation datasets}
The proposed model is first tested within in-domain BSP data: a clinical behavioral dataset from psychotherapy, in which conversations are characterized with clinically-relevant behavioral descriptors. 
Second, to evaluate the model's generalizability and domain robustness, we also test on  curated corpora of several ``out-of-domain" speech files, which contain diverse sources of speech from different scenarios such as comedy shows and debates.

\begin{table}[ht]

\centering
\caption{Description of behavior codes in Couples Therapy Corpus}
\begin{tabular}{c|l}
\hline
\multicolumn{1}{c|}{\begin{tabular}[c]{@{}c@{}}Behavior\\ Code\end{tabular}} & \multicolumn{1}{c}{Brief description} \\ \hline
 Acceptance     &  \makecell[l]{Indicates understanding, acceptance,\\ respect for partner’s views, feelings and behaviors }                              \\ \hline
 Blame          &    \makecell[l]{ Blames, accuses, criticizes partner, \\and uses critical sarcasm and character assassinations}                            \\ \hline
 
 Humor          &    \makecell[l]{ Includes jokingly making fun of self, lightly teasing the \\spouse, or making a reference to a mutually shared joke.}                            \\ \hline
 
Negativity       &   \makecell[l]{Overtly expresses rejection, defensiveness, \\blaming, and anger }                 \\ \hline
Positivity       &   \makecell[l]{Overtly expresses warmth, support, acceptance, \\affection, positive negotiation}  \\ \hline
\end{tabular}
\label{tab:beh_des}
\end{table}

\begin{table}[ht]
\caption{Description of evaluation data in Diverse Speech Behavior corpora}
\label{tab:EoData}
\centering
\begin{tabular}{c|c|l}
\hline
Category                          & ID & \multicolumn{1}{c}{Brief description}                                                                 \\ \hline
\multirow{2}{*}{Comedy show}      & 1  & George Carlin                                                                                   \\ \cline{2-3} 
                                  & 2  & Steve Hofstetter                                                                                \\ \hline
\multirow{2}{*}{Political debate} & 3  & Final Republican Presidential Debate, 2015                                                      \\ \cline{2-3} 
                                  & 4  & Vice Presidential Debate 2012                                                                   \\ \hline
\multirow{2}{*}{TED Talk}         & 5  & TEDtalk: Kevin Slavin                                                                         \\ \cline{2-3} 
                                  & 6  & TEDtalk: Christopher Steiner                                                                    \\ \hline
\multirow{2}{*}{Eulogy}           & 7  & Eulogy for a Son (youtube)                                                                      \\ \cline{2-3} 
                                  & 8  & \begin{tabular}[c]{@{}l@{}}Mr. Li Hongyi's Eulogy for the late \\ Mr. Lee Kuan Yew\end{tabular} \\ \hline
\end{tabular}
\end{table}

\subsubsection{Couple therapy dataset} \label{data:couple}
The first dataset we employ is the couples therapy corpus collected by the researchers in the UCLA/UW Couple Therapy Research Project \citep{christensen2004traditional}, in which 134 real couples were involved in a longitudinal study of 2 years for the evaluation of complex human behaviors related to marital therapy. 
In each session, a relationship-related topic (\eg ``Why cannot you leave my stuff alone?") was initiated and the couple had a conversation about this topic for 10 minutes. 

For evaluation purposes, we employ the annotation labels. In this couple therapy corpus, each participant's behaviors were evaluated based on the Couples Interaction \citep{heavey2002couples} and Social Support Rating Systems \citep{jones1998couples}.
The original 31 behavior codes were rated on a scale of 1-9, where 1 indicates the absence of the given behavior and 9 refers a strong presence. 
Similar to a previous study \citep{black2013toward}, we utilize five of the behaviors by binarizing the top and bottom 20\% of the original rating scores.
A brief description of the behavior codes used in this work is listed in Table \ref{tab:beh_des}.

\subsubsection{Curated speech data from different scenarios}
To further test the domain robustness of unsupervised behavior modeling method, we collected audio files representing a variety of other human spoken interaction domains. 
We manually collected audio files from two distinct speakers from four different scenarios: stand-up comedy routines, political debates, TED talks and eulogies. 
The audio names are listed in Table \ref{tab:EoData} and the duration of each audio is around 10 minutes.

\section{Experimental setup}
\label{sec:exp_setup}
\subsection{Audio data preparation}
For the training data, the audio files are directly extracted from movie video and combined into one single audio channel.
We do not perform any pre-processing procedures (\eg VAD and diarization) on the training data. Thus, the audio frames of movie can include conversations, silence, background music, and changing of speaker regions.

For couples therapy data, since each session consists of a dyadic conversation and the behavior ratings are provided for each spouse individually, we need to diarize the interactions to obtain the speech regions for each person.
We employ the pre-processing procedures described in the work \citep{black2013toward}. In short, we select sessions with an Signal-to-noise ratio (SNR) above 5dB, and conduct Voice Activity Detection (VAD) and Speaker diarization.
Speech regions from each session for the same speaker are used to analyze behaviors. 
The corpus has around 48 hours of audio data after these processing procedures.
More details of the data processing steps can be found in \citep{black2013toward}.

\subsection{Feature Extraction}
\label{subsec:feature_extraction}
We extract acoustic features, including speech prosody (pitch, intensity and their derivatives), spectral envelope characteristics (MFCCs, MFBs, LPCs and their derivatives), and voice quality (jitter, shimmer and their derivatives). 
The dimensions of MFCCs, MFBs and LPCs are 15, 8 and 8 respectively.
These Low-Level Descriptors (LLDs) are extracted using a 25 ms Hamming window with 10 ms shift. 
Within each analysis frame, we compute functionals of these acoustic features including Min (1st percentile), Max (99th percentile), Range (99th percentile – 1st percentile), Mean, Median, and Standard Deviation using openSMILE toolkit \citep{eyben2010opensmile}. 
These features are widely used and have shown effectiveness in many affect related tasks such as speech emotion recognition \citep{schuller2018speech}.

The size of analysis frame for target behaviors herein are larger than other shorter duration affective states (\eg of expressed emotions which can be reliably observed within a few seconds \citep{schuller2012avec}, one sentence \citep{zadeh2018multimodal} or a speaker turn \citep{busso2008iemocap}).
Previous behavioral annotation manuals \citep{heyman2004rapid,heavey2002couples} and computational analysis \citep{li2019linking}  report that the length of observation window for target behaviors is generally around 30 seconds or even longer. 
Based on these studies, in this work, in order to estimate meaningful behavioral metrics while maintaining a high resolution, the analysis frame size is set to 20 seconds with 1 second shift, the same as in previous works \citep{li2016sparsely, haoqibeh2vec, xia2015dynamic}. 
Under the feature configuration described above, for each analysis frame window, we have a feature dimension of 420. 

\subsection{Model configurations and parameter setting}
The training pairs are from movie audio, within a stationary region, the maximum sampling shift size $k$ is set to 6 seconds. 
For each frame $x_a$, we randomly select 4 context frames from neighboring segments as reconstruction frames $x_p$. 
While the frame pair ($x_n, x_{n\_p}$) is randomly selected from one stationary neighboring window in a different movie. 

In our experiment, the encoder-decoder structure of DCN and TE-DCN contains six hidden layers connected by PReLU \citep{he2015delving} activation function.
The dimension of the hidden layers are 300, 200, 64, 200, 300 respectively. 
The output of bottleneck embedding layer with 64 dimensions is regarded as behavior related representation that we are interested in. 
We use the Euclidean distance as the distance metric $D(\cdot,\cdot)$ in Equation (\ref{eq:triplet_loss}).
The model is trained with the Adam optimizer \citep{kingma2014adam} using a learning rate of 0.001 and a decay of 0.1 every 10 epochs. 
The triplet loss is optimized with a margin of $m$=2 and regularization weight of $\lambda$=0.01. 
We utilize different movie pairs as the validation set to terminate training with early stopping.

\subsection{Evaluation Method}
\label{subsec:eval_method}
\subsubsection{Evaluation method for in-domain Couples Therapy Corpus}
\label{subsubsec:eval_method_couple}
Considering the inter-annotator agreement, we binarize the original behavior ratings to model the evaluation task as a binary classification task of low- and high- presence of each behavior as in \citet{black2013toward}.
For each behavior code and each gender, we selected 70 sessions on one extreme of the code (e.g., high blame) and 70 sessions at the other extreme (e.g., low blame). 
This also enables balancing for each behavior resulting in classes of equal size.
As mentioned in section \ref{sec:dataset}, the couples therapy corpus only has session-level behavior code ratings. 
With these session-level labels, we evaluate the model in a supervised manner, though the behavior representation is trained in an unsupervised way with an out-of-domain movie corpus.

For each frame, once we obtain the latent behavioral manifold representation, we use the k-nearest neighbors algorithm to find a ``reference label". 
The value of k can be a hyper-parameter. To compare with existing work \citep{haoqibeh2vec}, similarly, we choose $k$=1 and use Euclidean distance to find the nearest frame among all remaining labeled frames from different sessions.
In addition, we also ensure that speaker characteristics information is not involved during testing by using leave-one-couple-out cross validation. 
Finally, majority voting is employed to generate session-level binary labels from multiple frame-level labels.

\subsubsection{Evaluation Method for Diverse Speech Behavior Corpora}
This evaluation is targeted to reflect different behavior or scenario styles. 
For example, as listed in Table \ref{tab:EoData}, the behavioral style from a stand-up comedy show is expected to be similar across performers, but expected to be different from those in a speech during a eulogy. 
Instead of focusing on scenario classification of whole speech regions, we are  interested in the level of similarity across different scenarios.
With this expectation, we calculate the results obtained by frame clustering with nearest neighbor, \ie which frame is close to which, as a percentage. 
This percentage score can be regarded as an indicator of style similarity among audio frames.

\section{Experimental results and discussions}
\label{sec:results}

\subsection{Experiment results of Couple Therapy Corpus}
The performance of couples' behavior classification results across different models is shown in Table \ref{tab:results_couple}. 
Besides the DCN and TE-DCN models, we further compare the results with four other models.

\begin{table}[ht]
\centering
\caption{Classification accuracy (\%) of behavior codes in Couple Therapy Corpus}
\label{tab:results_couple}
\scalebox{0.8}{
\begin{tabular}{ccccccc}
\hline
Behavior   & Baseline & DCN &  \begin{tabular}[c]{@{}c@{}}Triplet\\network \end{tabular}  & TE-auto-encoder & TE-DCN & \begin{tabular}[c]{@{}c@{}}Supervised training \\ in \citet{li2019linking}\end{tabular} \\ \hline
Acceptance & 57.14    & 66.43           &  60.71  &    65.71        & \textbf{68.21}  & 72.50                                                             \\ 
Blame      & 55.00    & 61.07            &  63.21  &    61.43        & \textbf{64.64}  & 71.79                                                             \\ 
Humor      & 54.29    & 55.00             &  56.79  &    \textbf{60.36}        & \textbf{60.36}  & -                                                    \\ 
Negativity & 63.92    & 63.93            &  61.79  &     60.71       & \textbf{66.43}  & 76.07                                                             \\ 
Positivity & 50.71    & 65.00            &  58.57 &      61.43      & \textbf{65.35}  & 65.36                                                             \\ \hline
Average    & 56.212   & 62.286          & 60.214   &    61.928  &  \textbf{64.998}    &    71.43 \\ \hline
\end{tabular}
}
\end{table}

\subsubsection{\textbf{Baseline model}}
For each behavior code, the number of behavior presence and absence sessions are balanced. 
Thus, a weak baseline of classification accuracy is 50\%. 
In this work, we use a better baseline model, which is built through the nearest neighbor classification in the original acoustic feature space.
Similarly, the session-level label is obtained by majority voting. 
The average classification accuracy of the five behavior codes is 56.212\%, which is slightly better than the weak baseline (random guess).
These results indicate that further representation learning process is necessary to extract behavior information from high dimensional acoustic features \citep{haoqibeh2vec}.

\subsubsection{\textbf{DCN model}}
In Table \ref{tab:results_couple}, for all behavior codes, the DCN model outperforms the baseline and achieves an average classification accuracy of 62.29\%. With the McNemaar test, compared with the baseline, the results are statistically significant with $p < 0.01$. Further details of the DCN model can be found in our previous work \citep{haoqibeh2vec}. 
These preliminary results support the possibility of using out-of-domain data for low-resource domain behavior modeling. 
Through it is not guaranteed that the extracted representations remove all other nuisance factors and only contain target behavior information, the results from the DCN model  show that affect related information are captured in the proposed manifold space. 

\subsubsection{\textbf{Triplet network model}}
As a comparison, we also perform the experiment with the triplet network model. 
Different from reconstruction of neighboring frames in DCN model, the triplet model only uses discriminative distance metric to directly optimize the representations within behavioral manifold.
Compared with the TE-DCN, the model does not contain the decoder parts. 
Thus, the contextual reconstruction loss from the decoder is not considered during the training and we only optimize the triplet loss from the outputs of encoders.

The experiment is conducted with similar settings as before, and we observe that the triplet model outperforms the baseline with average classification accuracy of 60.21\%. 
We notice that, for most behaviors, the DCN model achieves slightly better performance than this triplet model. 
In addition, we also tried negative sampling strategies \citep{schroff2015facenet,hermans2017defense} in the selection of triplet pair during training, however, we find that there is no improvement in terms of the domain data classification accuracy.

Considering the complexity of the training data, one reasonable explanation of the lower average performance of the triplet model might be the importance of the ``generative" property of decoder.
The representation of the behavioral manifold is trained to have the ability of encompassing and reconstructing the acoustic features of its neighbor frames, which are highly related to affect related information. 
The triplet network is only trained to discriminate samples with distance metric. 
Such a model might be failing to ensure that the optimized embeddings are highly relevant to capturing behavioral information, resulting a lower performance on the behavior modeling tasks.

\subsubsection{\textbf{TE-autoencoder model}}
Further, we test the TE-autoencoder model, a variant of the TE-DCN model. 
In the TE-autoencoder model, we replace the TE-DCN's contextual encoder-decoder structure with an autoencoder. 
Thus, once we have the training input pair $(x_a, x_p, x_n)$, the corresponding reconstruction pair is $(x_a, x_p, x_n)$ rather than previous context-based $(x_p, x_a, x_{n\_p})$.
The autoencoder is used to compress the original acoustic features and obtain representations with the same reduced dimension. 
Under this setting, the model can preserve the property of feature compression while ignoring the contextual information.
Similarly, this behavioral representation is optimized through both reconstruction loss and triplet loss in the target manifold.
We find that the results of the average performance of TE-autoencoder is worse than TE-DCN. 
This further supports the importance of contextual information, and also validates the behavior stationarity assumption.

\subsubsection{\textbf{TE-DCN model}}
The TE-DCN is built upon DCN, and the extracted behavior representation is enhanced by the discriminative metric under the behavior stationarity assumption.
From the classification results, we can observe that there is an improvement, from the 56.21\% of baseline to 64.99\% of TE-DCN model in terms of the averaged classification accuracy. 
Under the McNemar test, these results of proposed TE-DCN are statistically significant with $p<0.01$.
The TE-DCN model shows best performance across all models. In addition, compared with both DCN and triplet models, for all five behavior codes, a consistent improvement is obtained.

Moreover, we notice the complementary nature of DCN and triplet models in behavior modeling.
By combining these two, TE-DCN shows that both metric learning and context information can contribute to the overall unsupervised behavior modeling performance. 
These results are encouraging considering only unsupervised approaches are utilized with unlabeled out-of-domain data in TE-DCN. 

\subsubsection{\textbf{Supervised training method}}
The last column of the table indicates the classification results generated from a context-aware model via utilizing emotion related representation as behavioral primitives to facilitate the behavior quantification. 
Details of this supervised training approach can be found in \citep{li2019linking}.

These supervised classification results can be regarded as an upper bound performance of the supervised versus the unsupervised methods.
Moreover, it is necessary to mention that due to the complexity of human behavior and the subjectivity in annotation process, even for human annotators, the inter-annotator agreement can only reach about Krippendorff's $\alpha=0.8$ \citep{tseng2016couples}. 
Thus, although worse than the supervised method, the TE-DCN's performance is encouraging considering the fact that classification is obtained by a completely unsupervised method with simple majority vote.

\subsection{Behavioral trajectory analysis} 
In scenarios such as psychotherapy, instead of obtaining session-level classification labels, domain experts might be more interested in dynamic behavior change trajectories.
These trajectories can help the psychologists quickly locate the most salient regions and potentially reduce the workload of manual annotation. 
In this subsection, we use the couple therapy corpus as an example to illustrate that our unsupervised behavior modeling method can potentially provide such behavioral trajectories.

\begin{figure}[ht]
\centering
\includegraphics[width=0.6\textwidth]{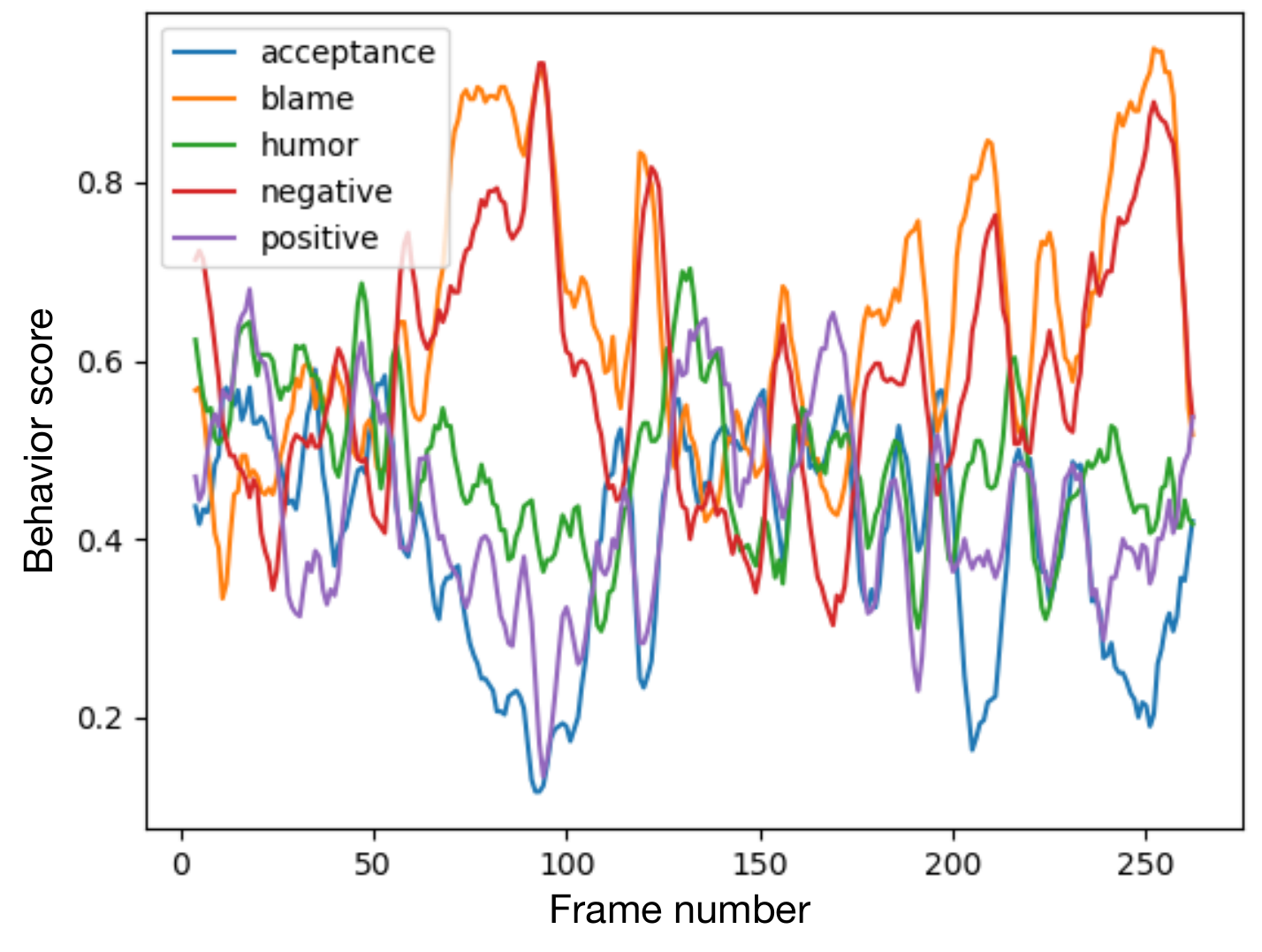}
\caption{One sample session with 5 behavior score trajectories}
\label{fig:fig_traj}
\end{figure}

Suppose we use the labeled frame samples as reference, and select the top $N$ nearest samples in the behavior manifold space. 
Among the top $N$ reference frames, we can calculate the percentage of samples labeled with the presence of a certain behavior code label (samples with label 1 in our case). 
For each test frame, the percentage value can indirectly imply the behavior ratings at some level. 
Figure \ref{fig:fig_traj} shows an example with one sample session's behavior dynamic change trajectories among five behaviors, and we set $N$ = 60 in this case.
In Table \ref{tab:ratings}, we provide the original averaged human annotation ratings and the automatically assigned behavior classification labels of this session.

\begin{table}[ht!]
\centering
\caption{Original annotation ratings and binarized classification labels for each behavior code}
\begin{tabular}{ccc}
\hline
Behavior   & \begin{tabular}[c]{@{}c@{}}Binarized label\\ (0:absence; 1: presence)\end{tabular} & \begin{tabular}[c]{@{}c@{}}Manual rating\\ (ranging from 1-9)\end{tabular} \\  \hline
Acceptance & 0                                                                               & 2.33                                                            \\ 
Blame      & 1                                                                               & 7.66                                                            \\ 
Humor      & 0                                                                               & 1.0                                                             \\ 
Negativity & 1                                                                               & 6.25                                                            \\ 
Positivity & 0                                                                               & 1.5                                                             \\ \hline
\end{tabular}
\label{tab:ratings}
\end{table}

Although the corpus does not provide utterance- or frame- level annotations, from this figure, we can notice the correlations among different predicted behavior code ratings.
We observe that behaviors Blame and Negativity are highly correlated, and behavior Positivity, Acceptance and Humor tend to have a similar trend. 
In addition, ``positive" related and ``negative" related behaviors have the opposite trend, which is in agreement with our intuition and previous supervised modeling research work \citep{black2013toward, li2016sparsely}. 
From the plot, we can also observe this session shows more presence of ``negative" behaviors (with higher scores) and less degree of ``positive" behaviors (with lower scores), which is in agreement with the human ratings listed in Table \ref{tab:ratings}.

In real world scenarios, it is often the case that the amount of annotated data might not be adequate to train a supervised behavior recognition system well. 
Through our unsupervised behavior modeling approach, if we need to annotate a newly collected session, this behavioral trajectory can quickly indicate salient behavior regions and help domain experts to locate and annotate the corresponding regions efficiently. 

\subsection{Experiment results on Diverse Speech Behavior Corpora}

\begin{figure}[h]
\centering
\includegraphics[width=0.55\textwidth]{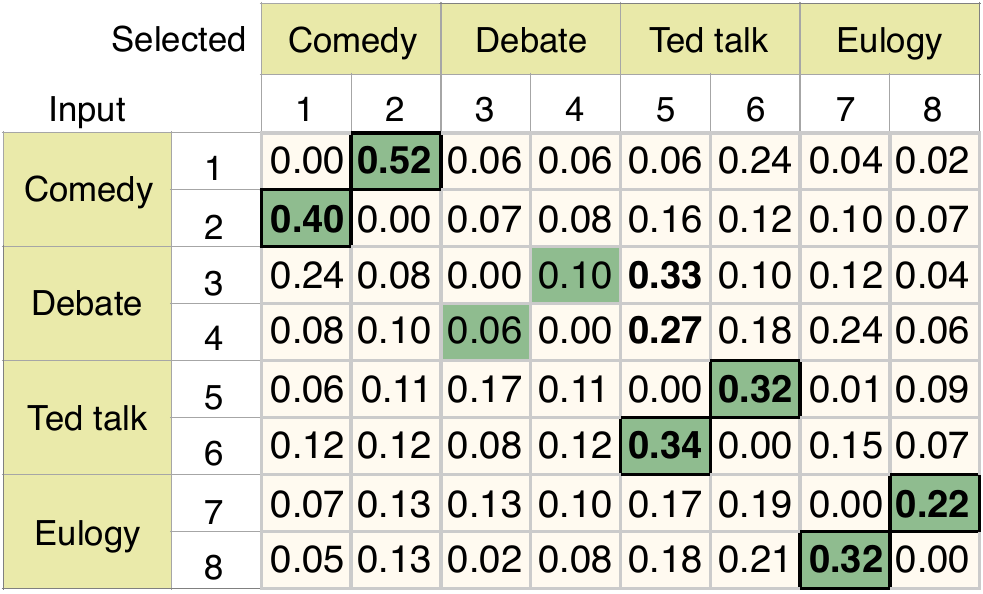}
\caption{Confusion matrix of behavior scenario similarity evaluation}
\label{fig:fig_ood}
\end{figure}

In this subsection, we use collected out-of-BSP domain data to evaluate the generalizability of TE-DCN model.
As listed in Table \ref{tab:EoData}, we collected two audio files from different speakers for each category.
The results of similarity evaluation among different scenarios is shown in Figure \ref{fig:fig_ood}. 
As described in Section \ref{subsec:eval_method}, in this table, each entry is calculated by dividing the number of nearest frames in each selected file by the total number of frames in the input audio. 
This normalized percentage value is used to evaluate the behavior similarity.
Ideally, audio from similar scenarios should exhibit high similarity with one another, and a lower score should be obtained within less related scenarios. 

From the similarity confusion matrix, we observe that for comedy, TED talk and eulogy categories, audio files exhibit high similarity scores within same category, and have lower scores for less related scenarios, as we expected. 
However, for the debate files, they are mostly confused with the ted talk files while also showing large similarity values with other scenarios. 
The reason for this might be the fact that during the debate, different politicians employed different kinds of debate skills and behaviors vary among different situations and topics.
In general, we find eight out of ten files are classified correctly based on the majority vote on frame-level clustering. 
Moreover, from the results table, we also can observe the similarity under different degrees among the different scenarios considered. 
All these results show promising behavioral quantification ability and potential applications of our proposed model.

\subsection{Nuisance factors and selection of features}
The TE-DCN model tends to preserve the shared behavioral information between the input frame and its neighboring frames.
We acknowledge that the design of the model combined with the behavioral stationarity assumption may have a potential complication: the neighboring frames could also encode speaker characteristics as well as acoustic conditions such as of the environment and channel conditions. 

To minimize the effect of these nuisance factors, the choice of input feature is critical in our proposed model. 
In addition to the triplet loss, the input features are designed to ensure that the unsupervised behavior modeling focuses more on affect related aspects rather than only employing the contextual information itself.
As described in section \ref{subsec:feature_extraction}, we directly use affect related hand-crafted features as input rather than extracting intermediate representations from raw spectrum features (\eg MFCC or MFB coefficients). 
We further replace the encoder-decoder structure of TE-DCN with CNN layers to input lower level raw spectrum features directly.
Based on the experiments, we find it is still challenging to extract behavioral representation exclusively, if inputs are lower level raw spectrum features, which largely contain other acoustically encoded information. 


\section{Conclusion}
\label{sec:conclusion}
The availability of adequate labelled data has been a critical bottleneck for supervised behavior modeling. Obtaining relevant behavioral data for such modeling often suffers from not only expensive data collection but varied and low human inter-annotation agreements. 
These constraints not only impact the modeling performance, but also limit the generalizability of the obtained behavioral models across domains.

In this work, we explore unsupervised learning for computational behavior modeling. 
We propose the TE-DCN model to extract behavioral representations in an unsupervised way.
The results suggest that the reconstruction with context information and metric learning are complementary methods within unsupervised behavior modeling. 
As a case study of unsupervised behavior modeling from speech using couples therapy data, our framework is shown to extract target behaviors from audio signals and achieves promising behavioral quantification results.
Although there is scope for improvement compared with the supervised method, our work provides possible solutions for the computational human behavior modeling: transfer information from out-of-domain data which are easily obtainable, and then adapt the model to specific domain applications.
We also note that information encoded in the speech unrelated to the target behaviors being modeled can negatively impact the representation learning. 

In the future, we plan to further computationally disentangle and reduce the speaker characteristics and other complex acoustic nuisance factors in the behavior representation. 
We plan to consider adversarial training to obtain more speaker-invariant and environment-robust behavior representations \citep{li2019speaker}.
Moreover, we also plan to investigate the feasibility of representation adaptation for downstream tasks by adding additional domain-specific supervised tuning.

\bibliographystyle{cas-model2-names}

\bibliography{cas-refs}


\end{document}